# Nonlinear Analysis of Chaotic Flow in a Three-Dimensional Closed-Loop Pulsating Heat Pipe

S. M. Pouryoussefi

Yuwen Zhang[1]

Fellow ASME

Department of Mechanical and Aerospace Engineering

University of Missouri

Columbia, MO 65211

**Abstract**

Numerical simulation has been conducted for the chaotic flow in a 3D closed-loop pulsating heat pipe (PHP). Heat flux and constant temperature boundary conditions were applied for evaporator and condenser sections, respectively. Water and ethanol were used as working fluids. Volume of Fluid (VOF) method has been employed for two-phase flow simulation. Spectral analysis of temperature time series was carried out using Power Spectrum Density (PSD) method. Existence of dominant peak in PSD diagram indicated periodic or quasi-periodic behavior in temperature oscillations at particular frequencies. Correlation dimension values for ethanol as working fluid was found to be higher than that for water under the same operating conditions. Similar range of Lyapunov exponent values for the PHP with water and ethanol as working fluids indicated strong dependency of Lyapunov exponent to the structure and dimensions of the PHP. An O-ring structure pattern was obtained for reconstructed 3D attractor at periodic or quasi-periodic behavior of temperature oscillations. Minimum thermal resistance of 0.85 ˚C/W and 0.88 ˚C/W were obtained for PHP with water and ethanol, respectively. Simulation results showed good agreement with experimental results from other work under the same operating conditions.

**Keywords**: Chaos, Numerical simulation, Pulsating heat pipe, Spectral analysis, Two-phase flow

1. Introduction

The Oscillating or Pulsating Heat Pipe (OPH or PHP) is a very promising heat transfer device. In addition to its excellent heat transfer performance, it has a simple structure. Unlike the conventional heat pipes, there is no wick structure to return the condensed working fluid back to the evaporator section. The PHP is made from a long, continuous capillary tube bent into many turns. The diameter of the PHP must be sufficiently small so that vapor plugs can be formed by capillary action. Due to the pulsation of the working fluid in the axial direction of the tube, heat is transported from the evaporator section to the condenser section. The heat input, which is the driving force, increases the pressure of the vapor plug in the evaporator section. In turn, this pressure increase will push the neighboring vapor plugs and liquid slugs toward the condenser, which is at a lower pressure [1]. Many experimental and theoretical studies have been conducted to understand complicated behaviors of PHP. Shafii et al. [1, 2] presented analytical models of thermal behavior and heat transfer for both open and closed-loop PHPs with multiple liquid slugs and vapor plugs. The results showed that gravity does not have significant effect on the performance of the unlooped PHPs with top heat mode. In addition, higher surface tension resulted in a slight increase in total heat transfer.

---

[1] Corresponding Author. Email: zhangyu@missouri.edu

Zhang et al. [3] investigated heat transfer process in evaporator and condenser sections of the PHP. They developed heat transfer models in the evaporator and condenser sections of a pulsating heat pipe with one open end by analyzing thin film evaporation and condensation. Zhang and Faghri analyzed oscillatory flow in pulsating heat pipes with arbitrary numbers of turns [4]. The results showed that the increase in the number of turns has no effect on the amplitude and circular frequency of oscillation when the number of turns is less than or equal to five. Zhang and Faghri [5] reviewed advances and unsolved issues in pulsating heat pipes. Shao and Zhang [6] studied thermally-induced oscillatory flow and heat transfer in a U-shaped minichannel. The sensible heat transfer coefficient between the liquid slug and the minichannel wall was obtained by analytical solution for a laminar liquid flow and by empirical correlations for a turbulent liquid flow. Kim et al. [7] analyzed entropy generation for a pulsating heat pipe. It was observed that the entropy generation is significantly affected by the initial temperature in the PHP. They also studied the effects of fluctuations of heating and cooling section temperatures on performance of a PHP [8]. They found that both amplitude and frequency of the periodic component of the temperature fluctuation affected the liquid slug displacements, temperatures and pressures of the two vapor plugs, as well as the latent and sensible heat transfer.

A mathematical model predicting the fluid motion and temperature drop in an OHP has been developed by Ma et al. [9]. The model included the forced convection heat transfer due to the oscillating motions, the confined evaporating heat transfer in the evaporating section, and thin film condensation heat transfer in the condensing section. An experimental investigation of temperature drops occurring in an OHP was also conducted. Researchers have investigated pulsating heat pipes from different aspects because of their importance and useful applications [10-13]. Qu et al. [14] conducted an experimental study on thermal performance of a silicon-based micro pulsating heat pipe. The effects of gravity, filling ratio, and working fluids on the overall thermal resistance were discussed. Experimental results showed that a micro-PHP embedded in a semiconductor chip could significantly decrease the maximum localized temperature [15]. Turkyilmazoglu investigated effects of nanofluids on heat transfer enhancement in single- and multi-phase flows [16-18]. It was shown that by increasing the diffusion parameter in the multi-phase model, more enhancements in the rate of heat transfer could be obtained. In addition, a rescaling method to simplify the evaluation of flow and physical parameters, such as skin friction and heat transfer rate in single phase nanofluids, was proposed. Xian et al. [19] experimentally investigated dynamic fluid flow in oscillating heat pipe under pulse heating. They employed a high-speed camera to conduct experiments and fluid flow visualization in PHP. Jiaqiang et al. [20] studied pressure distribution and flow characteristics of a closed oscillating heat pipe under different operating conditions. They analyzed the relationship of flow pattern distribution and pressure distribution.

Chaotic behaviors in PHPs have been reported in some recent experimental studies [21]. Dobson [22] theoretically and experimentally investigated an open oscillatory heat pipe including gravity. Convective heat transfer to and from both the vapor bubble and the liquid plug was not included in the theoretical model. A special open oscillatory heat pipe was manufactured for the experiment. It was found that the theoretical model is able to reflect the characteristic chaotic behavior of the experimental devices. Xiao-Ping and Cui [23] studied the dynamic properties for the micro-channel phase change heat transfer system by theoretical method combined with experiment. A dynamic model for micro-channel phase change heat transfer system has been established by considering disjoining pressure at the liquid– vapor interface due to mini size of channel. Power spectrum density analysis showed that the system is in a state of chaos. They concluded that chaos of the system is one of the important conditions that lead to high heat transfer performance for microchannel system by considering the correspondence of chaos with turbulence. Song and Xu [24] have run series of experiments to explore chaotic behavior of PHPs. FC-72 and deionized water were used as the working fluids. A high speed data acquisition system was used to record time series of temperatures at different locations on the PHP. FC-72 PHPs had complex relationship between correlation dimensions and number of turns. Correlation dimensions were increased by increasing the number of turns for water PHPs. Qu et al. [25] employed ethanol as working fluid to investigate the chaotic behavior of wall temperature oscillations in a closed-loop pulsating heat pipe. The experimental results were analyzed by using nonlinear analyses of the pseudo-phase-plane trajectories, the correlation dimension, the largest Lyapunov exponent and the recurrence plots. Chaotic states were observed in the recurrence plots of the temperature oscillations.

Although several theoretical and experimental studies of the chaotic behavior of pulsating heat pipes have been carried out, there has been no detailed numerical simulation for PHPs. Pouryoussefi and Zhang [26] recently conducted a numerical simulation of the chaotic flow in the closed-loop pulsating heat pipe. They considered a two dimensional structure for the PHP and water was the only working fluid. Constant temperature was the boundary condition for both evaporator and condenser. Chaotic behavior of the PHP was investigated under different operating conditions. In this paper, the thermal and chaotic behaviors of a three-dimensional PHP with more number of turns, using ethanol in addition to water as the working fluids, will be investigated. Heat flux and constant temperature boundary conditions have been used for evaporator and condenser, respectively. Chaotic parameters and thermal performance have been investigated under different operating conditions.

2. **Physical Modeling**

A schematic cross sectional view of the 3D PHP is shown in Fig. 1. In order to validate the simulation results of current study with an experimental test case, the structure and dimensions of the pulsating heat pipe in simulation were considered the same as experimental investigation in [13]. The diameter of the tube is 1.8 mm. The lengths of the evaporator, adiabatic and condenser sections are 60 mm, 150 mm and 60 mm, repectively. Water and ethanol were used as working fluids. Different evaporator heating powers and filling ratios were tested for the numerical simulation. There are 25 points in Fig. 1 highlited with bold rectangles and these poits are assigned for tmeperature measurement during numerical simulation.

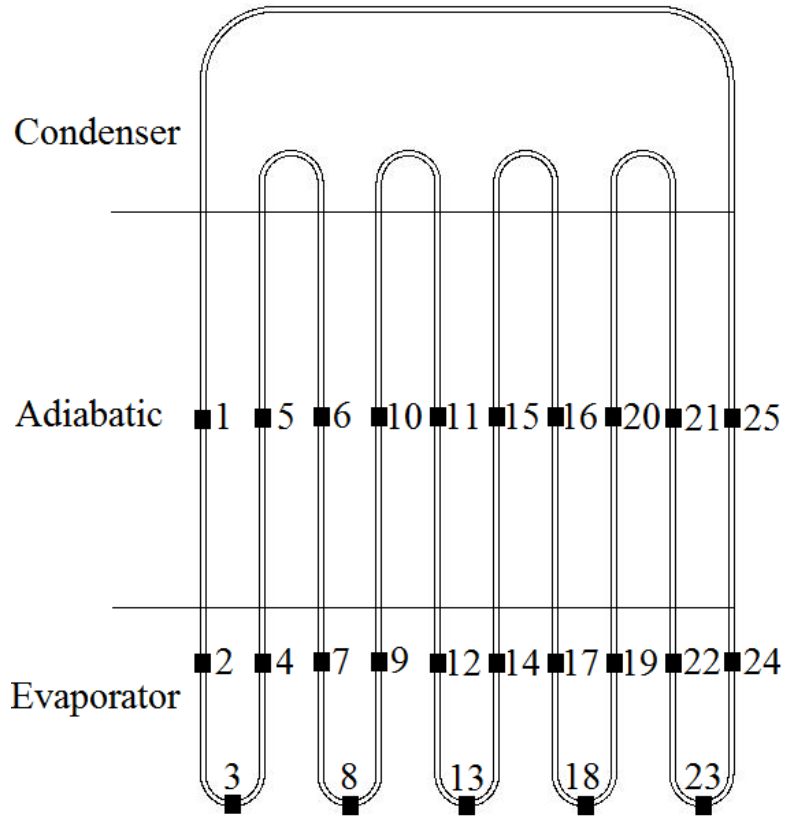

Figure 1 Schematic cross sectional view of the 3D PHP

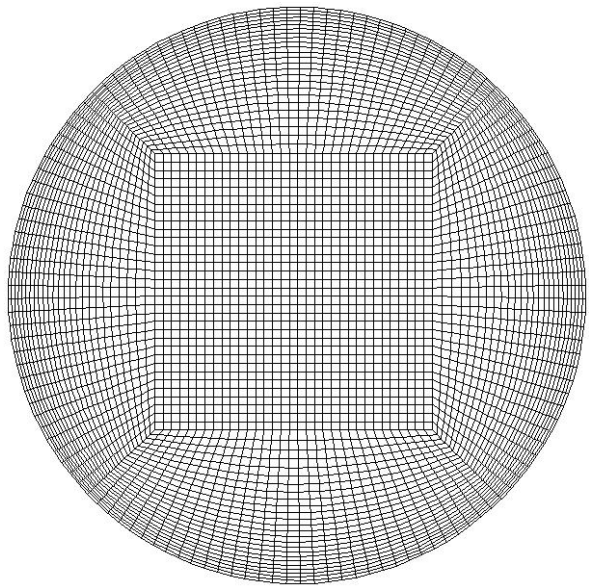

Figure 2 Meshing configuration

Figure 2 shows cross sectional profile of meshing configuration used in the numerical simulation. This profile has been sweeped through the entire centerline of the PHP led to create hexahedral meshes. Surface tension is an important factor in the performance of the pulsating heat pipe. Employing quadrilaterals or hexahedrals meshes for numerical simulation leads to more accurate results in case of surface tension than those of triangular and tetrahedral. The hexahedral mesh was used for the entire pulsating heat pipe. It is evident in Fig. 2 that number of layers are increased near the wall of the PHP to properly capture the flow gradients.

Volume of Fluid (VOF) method has been applied for two phase flow simulation. The volume of fluid method is a free-surface modelling technique, i.e. a numerical technique for tracking and locating the free surface (or fluid-fluid interface). It belongs to the class of Eulerian methods which are characterized by a mesh that is either stationary or is moving in a certain prescribed manner to accommodate the evolving shape of the interface. VOF is an advection scheme and numerical recipe that allows tracking the shape and position of the interface, but it is not a standalone flow solving algorithm. The Navier–Stokes equations describing the motion of the flow have to be solved separately. The same applies for all other advection algorithms. Applications of the VOF model include stratified flows, free-surface flows, filling, sloshing, the motion of large bubbles in a liquid, the motion of liquid after a dam break, the prediction of jet breakup (surface tension), and the steady or transient tracking of any liquid-gas interface [21].

Tracking of the interfaces between the phases is accomplished by the solution of a continuity equation for the volume fraction of one (or more) of the phases. For the $q^{th}$ phase, this equation has the following form [26]:

$$\frac{1}{\rho_q}\left[\frac{\partial}{\partial t}(\alpha_q \rho_q) + \nabla \cdot (\alpha_q \rho_q \mathbf{v}) = \sum_{p=1}^{n}(\dot{m}_{pq} - \dot{m}_{qp})\right] \qquad (1)$$

where $\dot{m}_{qp}$ is the mass transfer from phase q to phase p, and $\dot{m}_{pq}$ is the mass transfer from phase p to phase q due to phase change. The primary-phase volume fraction will be computed based on the following constraint:

$$\sum_{q=1}^{n} \alpha_q = 1 \qquad (2)$$

The volume fraction equation was solved using explicit time discretization. In the explicit approach, finite-difference interpolation schemes are applied to the volume fractions that were computed at the previous time step [26]:

$$\frac{\alpha_q^{n+1} \rho_q^{n+1} - \alpha_q^n \rho_q^n}{\Delta t} V + \sum_f (\rho_q U_f^n \alpha_{q,f}^n) = \left[\sum_{p=1}^{n}(\dot{m}_{pq} - \dot{m}_{qp}) + S_{\alpha_q}\right] V \qquad (3)$$

where n+1 is the index for new (current) time step, n is the index for previous time step, $\alpha_{q,f}$ is the face value of the $q^{th}$ volume fraction, V is the volume of cell and $U_f$ is the volume flux through the face, based on normal velocity.

The properties appearing in the transport equations are determined by the presence of the component phases in each control volume. In the vapor-liquid two-phase system, the density and viscosity in each cell are given by [26]:

$$\rho = \alpha_v \rho_v + (1 - \alpha_v)\rho_l \qquad (4)$$

$$\mu = \alpha_v \mu_v + (1 - \alpha_v)\mu_l \tag{5}$$

One momentum equation is solved throughout the domain, and the resulting velocity field is shared among all phases. The momentum equation, shown below, is dependent on the volume fractions of all phases through the properties ρ and μ [26]:

$$\frac{\partial}{\partial t}(\rho \vec{v}) + \nabla \cdot (\rho \boldsymbol{vv}) = -\nabla p + \nabla \cdot [\mu(\nabla \boldsymbol{v} + \nabla \boldsymbol{v}^T)] + \rho \boldsymbol{g} + \boldsymbol{F} \tag{6}$$

One limitation of the shared-fields approximation is that in cases where large velocity differences exist between the phases, the accuracy of the velocities computed near the interface can be adversely affected.

The energy equation, also shared among the phases, is [26]:

$$\frac{\partial}{\partial t}(\rho E) + \nabla \cdot \big(\boldsymbol{v}(\rho E + p)\big) = \nabla \cdot \big(k_{eff}\nabla T\big) + S_h \tag{7}$$

The VOF model treats energy, E, and temperature, T, as mass-averaged variables [26]:

$$E = \frac{\sum_{q=1}^{n} \alpha_q \rho_q E_q}{\sum_{q=1}^{n} \alpha_q \rho_q} \tag{8}$$

where $E_q$ for each phase is based on the specific heat of that phase and the shared temperature. The properties ρ and $k_{eff}$ (effective thermal conductivity) are shared by the phases and the source term, $S_h$ is equal to zero.

The surface curvature is computed from local gradients in the surface normal at the interface. The surface normal $\boldsymbol{n}$, defined as the gradient of $\alpha_q$, the volume fraction of the $q^{th}$ phase is:

$$\boldsymbol{n} = \nabla \alpha_q \tag{9}$$

The curvature, K, is defined in terms of the divergence of the unit normal, $\hat{\boldsymbol{n}}$ :

$$\boldsymbol{K} = \nabla \cdot \hat{\boldsymbol{n}} \tag{10}$$

where

$$\hat{\boldsymbol{n}} = \frac{\boldsymbol{n}}{|\boldsymbol{n}|} \tag{11}$$

is the unit surface normal vector. The surface tension can be written in terms of the pressure jump across the surface. The force at the surface can be expressed as a volume force using the divergence theorem. It is this volume force that is the source term which is added to the momentum equation. It has the following form [26]:

$$F_{vol} = \sigma_{12} \frac{\rho k_1 \nabla \alpha_2}{\frac{1}{2}(\rho_1 + \rho_2)} \tag{12}$$

where ρ is the volume-averaged density computed using Eq. (4). Equation (12) shows that the surface tension source term for a cell is proportional to the average density in the cell.

To model the wall adhesion angle in conjunction with the surface tension model, the contact angle that the fluid is assumed to make with the wall is used to adjust the surface normal in cells near the wall. This so-called dynamic boundary condition results in the adjustment of the curvature of the surface near the wall. If $\theta_w$ is the contact angle at the wall, then the surface normal at the live cell next to the wall is [26]:

$$\hat{n} = \hat{n}_w cos\theta_w + \hat{t}_w sin\theta_w \qquad (13)$$

where $\hat{n}_w$ and $\hat{t}_w$ are the unit vectors normal and tangential to the wall, respectively.

In this study, the heating power range for the evaporator is from 10 W to 120 W. Condenser has a constant temperature of 20 °C for all operating conditions. In addition, initial temperatue of the liquid is 27 °C for all operating conditions.

## 3. Results and Discussions

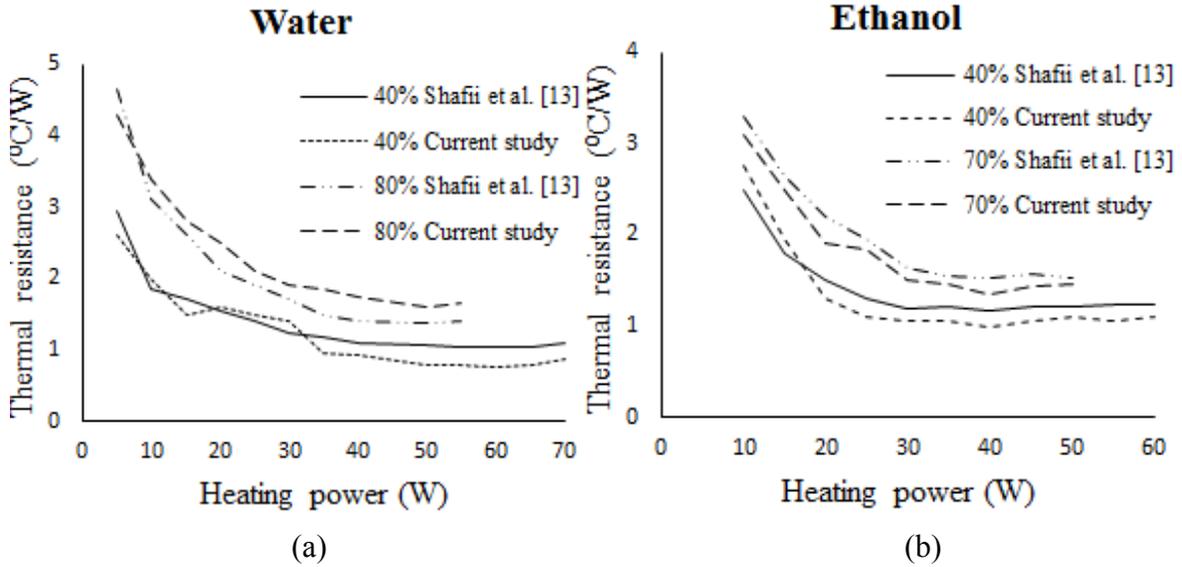

(a) (b)

Figure 3 Comparison of thermal resistance versus heating power between simulation results of current study and experimental results of reference [13]

As mentioned earlier, in order to validate the simulation results of current study with an experimental test case, the structure and dimensions of the pulsating heat pipe in simulation were considered the same as experimental investigation in [13]. In addition, the same working fluids of water and ethanol were employed. It should be noted that due to the experiment limitations as the mentioned investigation, the maximum heating power of 70 W and 60 W were applied for water and ethanol as working fluids respectively in [13]. Besides, constant temperature of 20 °C was used for the condenser boundary condition in the experiment. Figure 3 shows comparison of thermal resistance with respect to heating power between simulation results of current study and experimental results of Shafii et al. [13]. Thermal resistance behavior has been illustrated at filling ratios of 40% and 80% for water as working fluid and 40% and 70% for ethanol as working fluid. As seen in Fig. 3(a) the thermal resistance obtained from simulation has higher values after heating power of 10 W comparing with experimental results for water as working fluid and filling ratio of 80%. The maximum error occurs at heating power of 40 W which is almost 13%. The average error is 8.7 % under these operating conditions. At filling ratio of 40%, the maximum error occurs at heating power of 60 W which is almost 15%. The average error is 9.7% under these operating conditions. Figure 3(b) shows that the thermal resistance obtained from simulation has lower values for all heating powers comparing with experimental results for ethanol as working fluid and filling ratio of 70%. The maximum error occurs at heating power of 20 W which is almost 10%. The average error is 6.3 % under these operating conditions. At filling ratio of 40%, the maximum error occurs at heating power of 25 W which is almost 13%. The average error is 7.5% under these operating conditions. Figure 3 confirms there is a good agreement between simulation and experimental results. The range of errors percentage is normal for such comparison between simulation and experimental results. There are several source of error in any experiment such as instrument and measurement inaccuracy and uncertainty. One of the most important issues in this particular test case is insulation of evaporator and adiabatic sections which cannot be ideal experimentally. Thus, ideal insulation in numerical simulation is one of the significant reasons which causes differences between simulation and experimental results.

### 3.1. Volume Fractions

Figures 4 to 8 illustrate volume fractions of liquid and vapor at different times (red color represents the vapor and blue color represents the liquid) under different operating conditions. Figure 4(a) shows almost the initial condition of the PHP with ethanol as working fluid and filling ratio of 40%. Figure 5 demonstrates formation of vapor bubbles and fluid flow development in the PHP. Figure 6 depicts the volume fractions of liquid and vapor in the PHP after the fluid flow has been established. Mostly a similar process occurs in the PHP with water as working fluid and filling ratio of 65% (Figures 4(b), 7 and 8). One of the most significant effects due to increasing filling ratio is pressure increase in the PHP which leads to a longer time duration for boiling to start (because of increasing the saturated temperature) and slower flow motion in the PHP. In addition to higher filling ratio, due to higher specific heat and higher saturation temperature for water comparing to ethanol, boiling process starts later for the PHP with water as working fluid than PHP with ethanol as working fluid. The effect of change in filling ratio on the chaotic and thermal behavior of the PHP will be investigated in the next subsections afterwards. It is evident that liquid plugs having menisci on the plug edges are formed due to surface tension forces. A liquid thin film also exists surrounding the vapor plug. The angle of contact of the menisci, the liquid thin film stability and its thickness depends on the fluid-solid combination and the operating parameters which are selected. If a liquid plug is moving or tends to move in a specific direction then the leading contact angle (advancing) and the lagging contact angle (receding) will be different [21]. This happens because the leading edge of the plug moves on a dry surface (depending on the liquid thin film stability and existence) while the lagging edge moves on the just wetted surface. These characteristics can be observed in Fig. 9, which shows a snapshot of liquid-vapor plugs as result of simulation in the vertical part of the system.

The results showed that the fluid flow finally circulates in one direction (clockwise or counterclockwise) in the pulsating heat pipe. This direction is based on a random process and could be different even under the same operating and boundary conditions. The liquid and vapor plugs are subjected to pressure forces from the adjoining plugs. Forces acting on the liquid plugs are due to the capillary pressures created by the menisci curvatures of the adjacent vapor bubbles. The liquid plugs and vapor bubbles experience internal viscous dissipation as well as wall shear stress as they move in the PHP tube [21]. The relative magnitude of these forces will decide the predominant force to be considered. The liquid and vapor plugs may receive heat, reject heat or move without any external heat transfer, depending on their location in the evaporator, condenser or the adiabatic section. Probability of events frequently places vapor bubbles in direct contact with the evaporator tube surface. In this case saturated vapor bubbles receive heat which is simultaneously followed up by evaporation mass transfer from the adjoining liquid plugs thereby increase the instantaneous local saturation pressure and temperature.

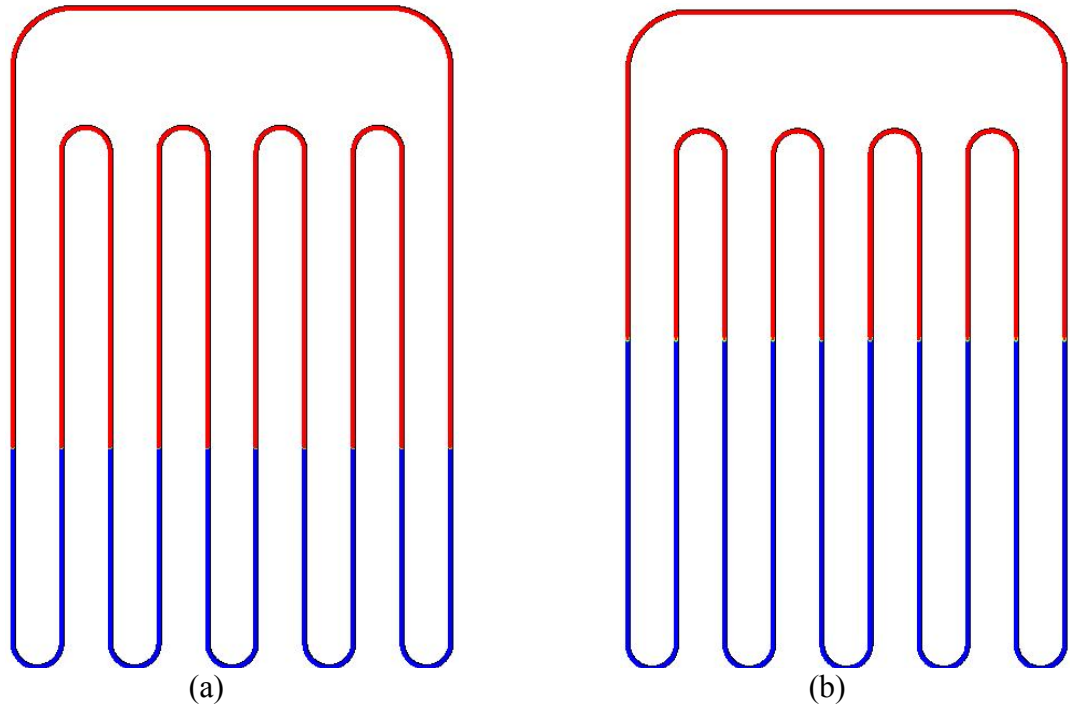

(a)                      (b)

Figure 4 Volume fractions of liquid and vapor at t=0.1 s under heating power of 70 W and condenser temperature of 20 ˚C for ethanol, FR of 40% (a) and water, FR of 65% (b)

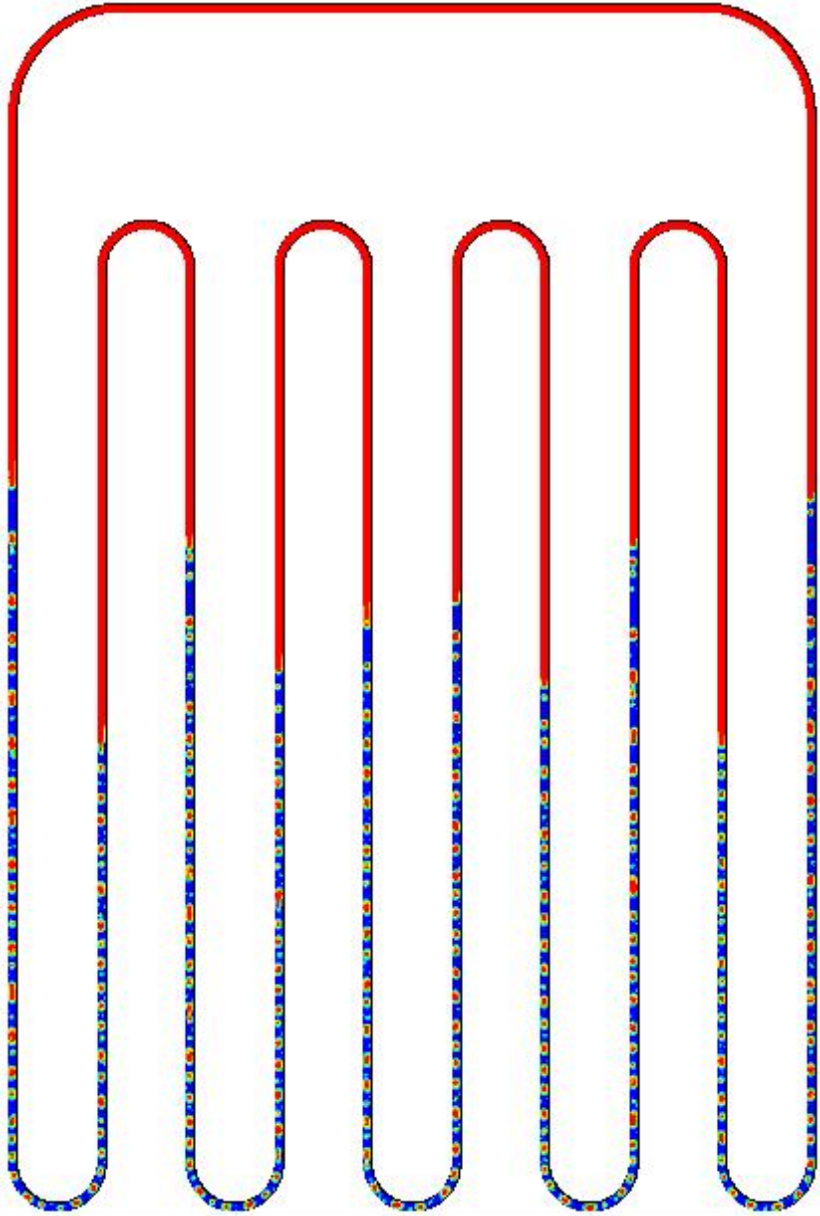

Figure 5 Volume fractions of liquid and vapor under heating power of 70 W and condenser temperature of 20 ˚C for ethanol as working fluid with FR of 40% at t=0.8 s

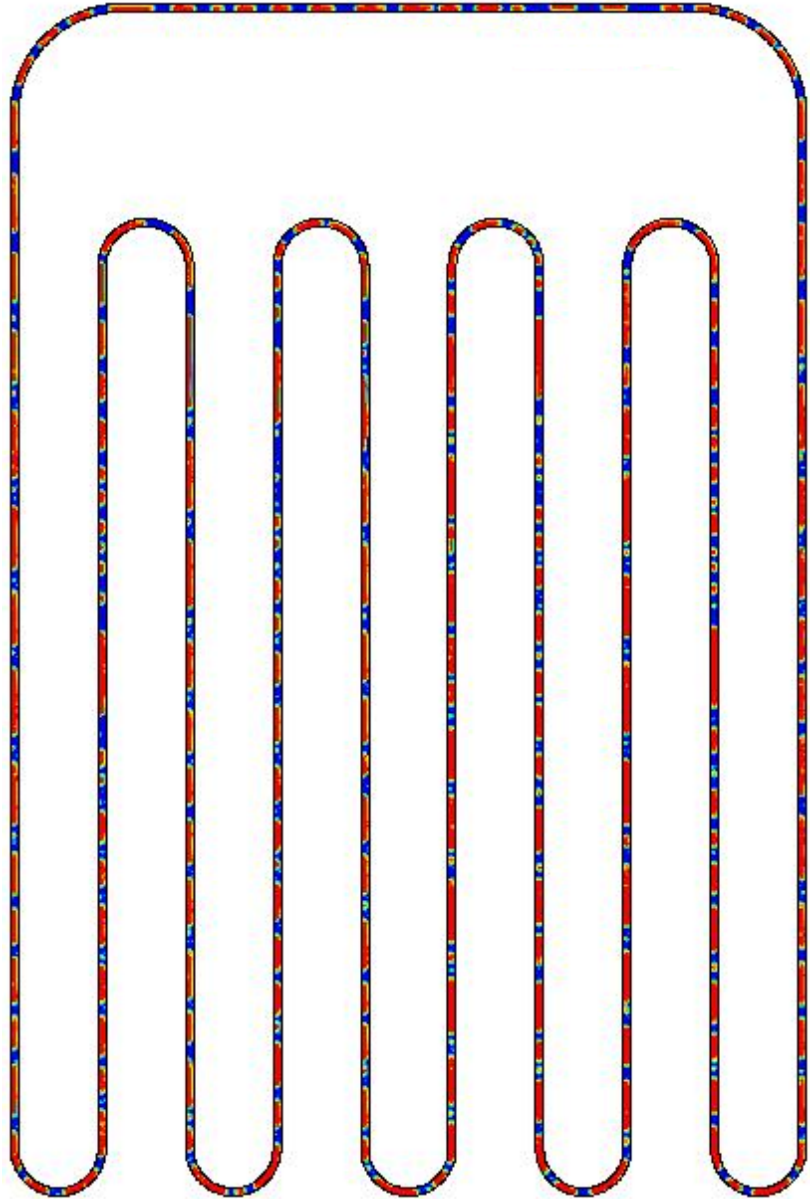

Figure 6 Volume fractions of liquid and vapor under heating power of 70 W and condenser temperature of 20 ˚C for ethanol as working fluid with FR of 40% at t=20 s

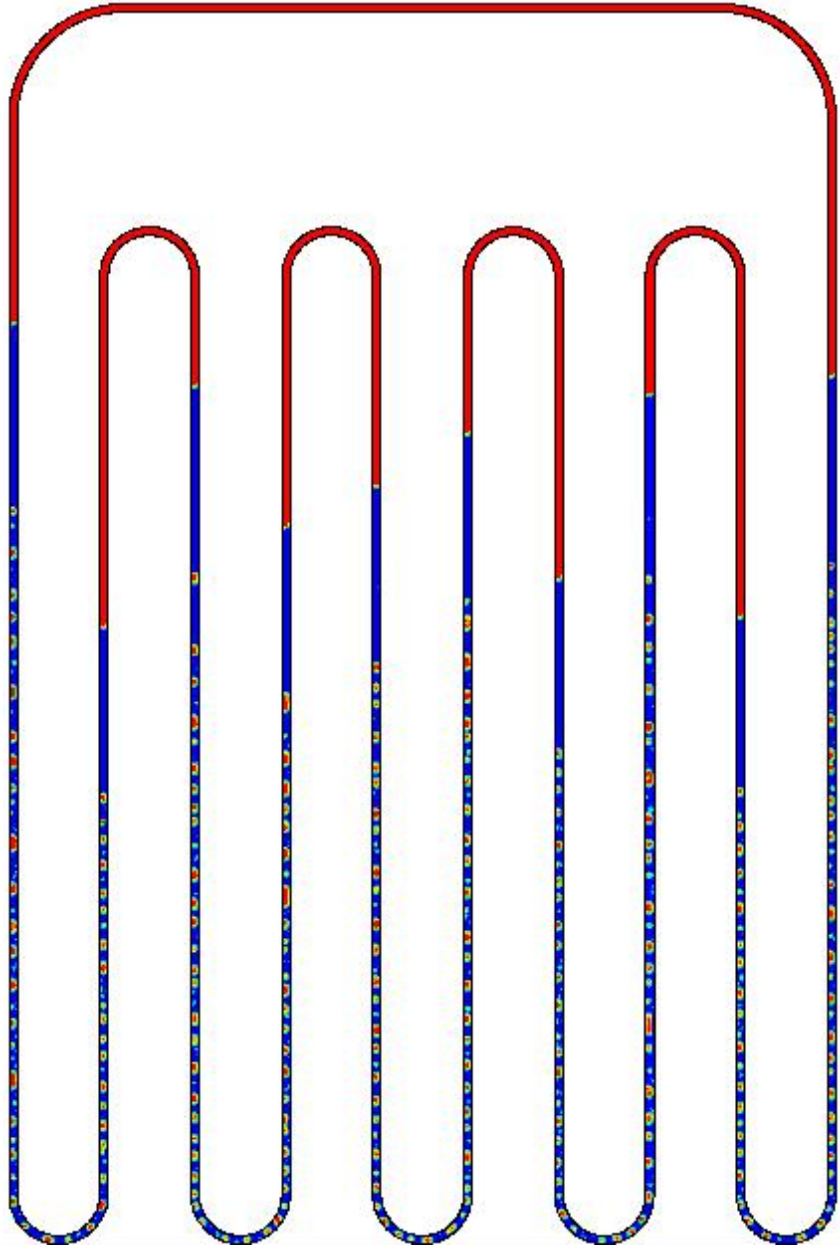

Figure 7 Volume fractions of liquid and vapor under heating power of 70 W and condenser temperature of 20 ˚C for water as working fluid with FR of 65% at t=0.8 s

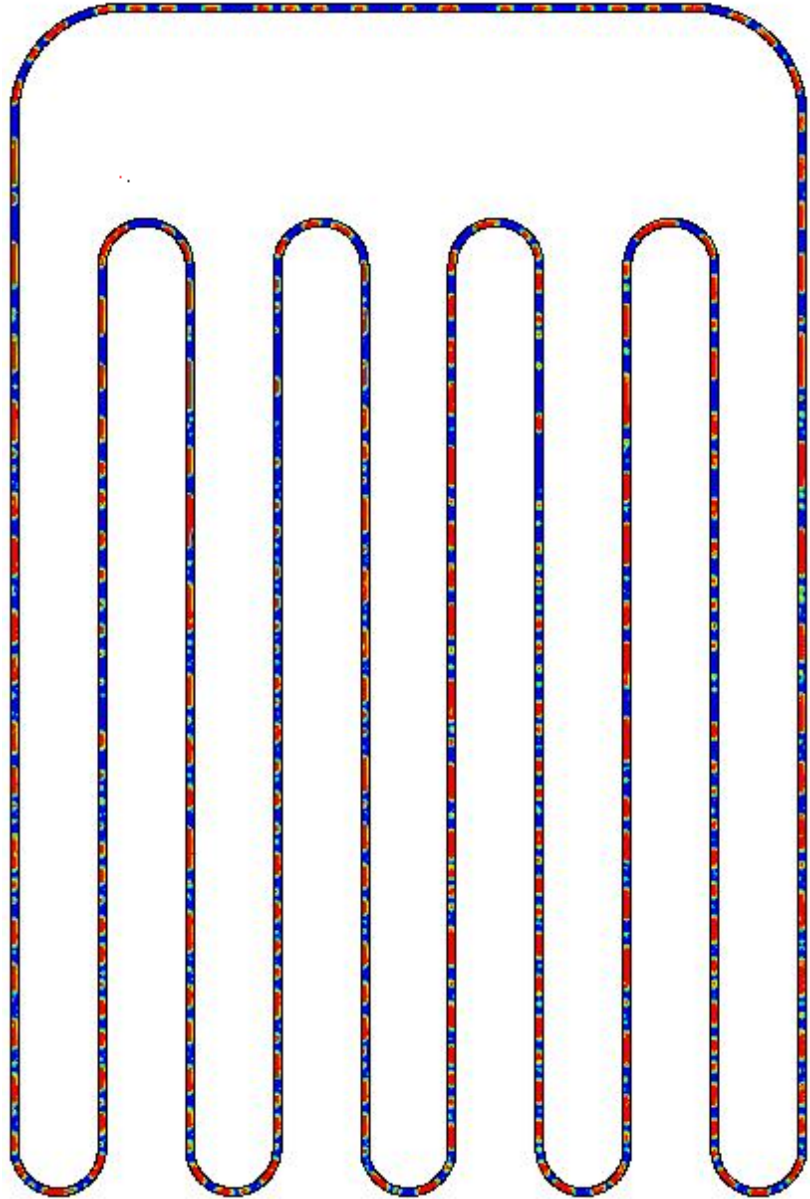

Figure 8 Volume fractions of liquid and vapor under heating power of 70 W and condenser temperature of 20 ˚C for water as working fluid with FR of 65% at t=20 s

Liquid thin film exists around the vapor plug, and the thickness of the liquid film may vary depending on some boundary conditions. One of the most important parameters affecting the liquid film is the wall temperature in the PHP. Since evaporator, condenser and adiabatic sections have different working temperatures, liquid film with different thicknesses may form in these regions [26]. Figure 10 shows different liquid films surrounding the vapor plugs with different thicknesses as result of simulation. Figure 10 illustrates vapor plugs in the evaporator (a), adiabatic section (b) and condenser (c) respectively. It can be seen that by decreasing the wall temperature from evaporator to condenser, the liquid film thickness increases.

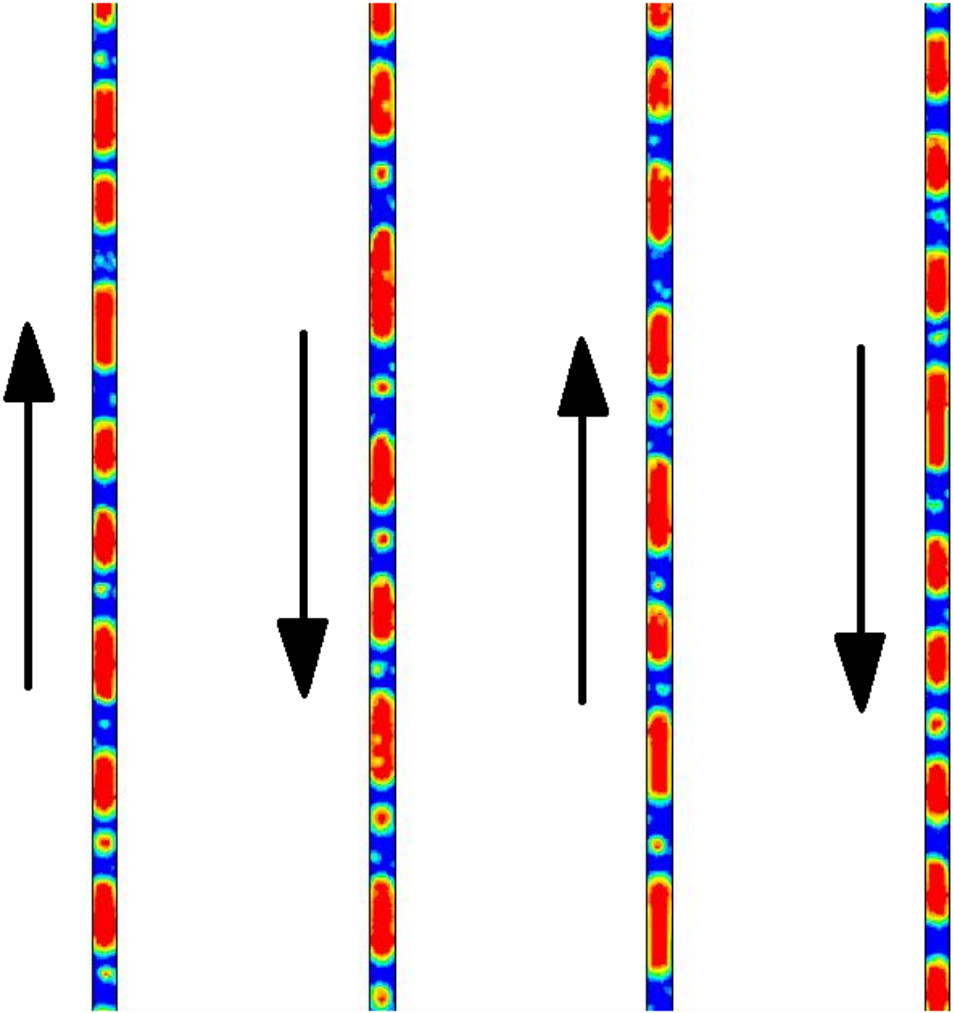

Figure 9 Vapor and liquid plugs in the PHP

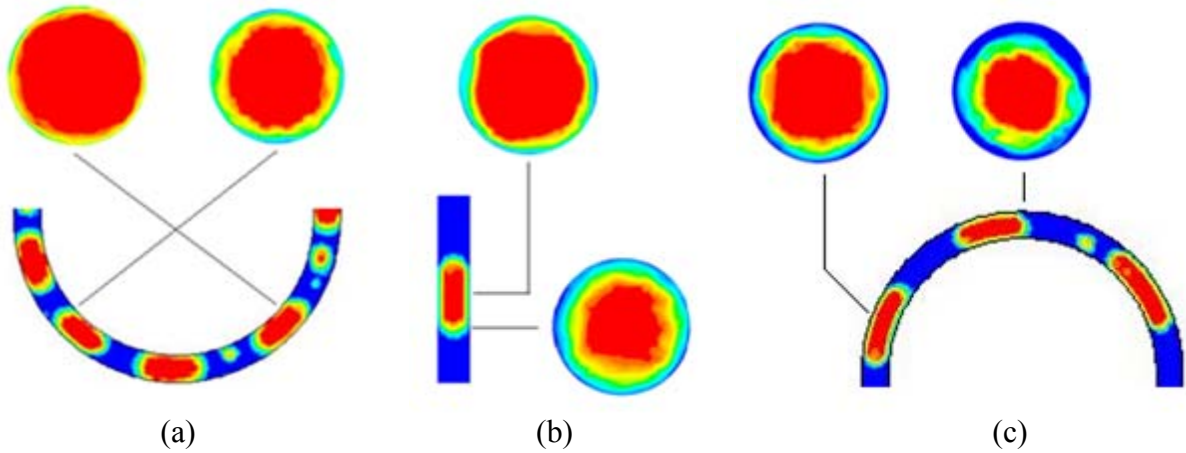

|  (a)  |  (b)  |  (c)  |

Figure 10 Liquid film around the vapor plugs at evaporator (a), adiabatic section (b) and condenser (c)

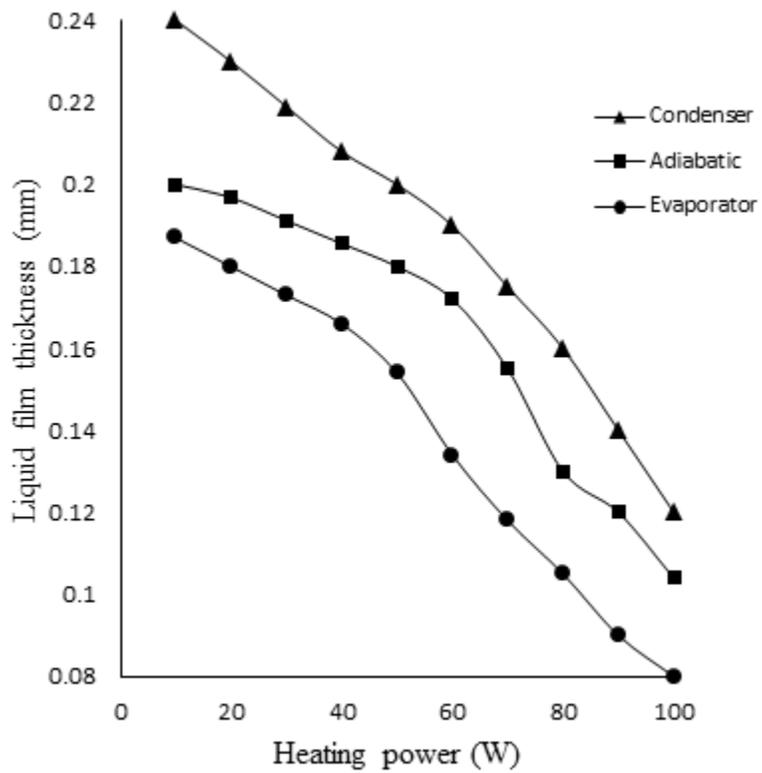

Figure 11 Liquid film thickness versus heating power

Figure 11 shows the comparison of the averaged thickness of liquid film for the vapor plugs at evaporator, adiabatic section and condenser under different heating power with condenser temperature of 20 °C and water as working fluid with filling ratio of 65%. It is evident that condenser has thicker liquid film than the adiabatic section and evaporator at each heating power and evaporator has thinner liquid films. The thickness decreases linearly from 10 W to 40 W in evaporator. By increasing the heating power higher than 40 W, a decrease in slope can be observed from 40 W to 100 W. A similar behavior for the thickness in adiabatic section is visible. But the change in slope occurs at a higher heating power of 60 W. The thickness in condenser decreases from 0.24 mm to 0.12 mm linearly with a constant slop by increasing the heating power.

In adiabatic section, while passing from the evaporator to the condenser, the train of vapor-liquid plugs is subjected to a series of complex heat and mass transfer process. Essentially non equilibrium conditions exist whereby the high pressure, high temperature saturated liquid-vapor plugs is brought down to low pressure, low temperature saturated conditions existing in the condenser. Internal enthalpy balancing in the form of latent heat takes place by evaporation mass transfer from the liquid to the vapor plugs whereby saturation conditions are always imposed on the system during the bulk transient in the adiabatic section [21]. Sometimes two vapor bubbles combine together to form a larger vapor slug [26]. Such a combination process in evaporator and adiabatic section is illustrated in Fig. 12 as a result of simulation; this phenomenon mostly was observed in the adiabatic section and was observed rarely in the condenser.

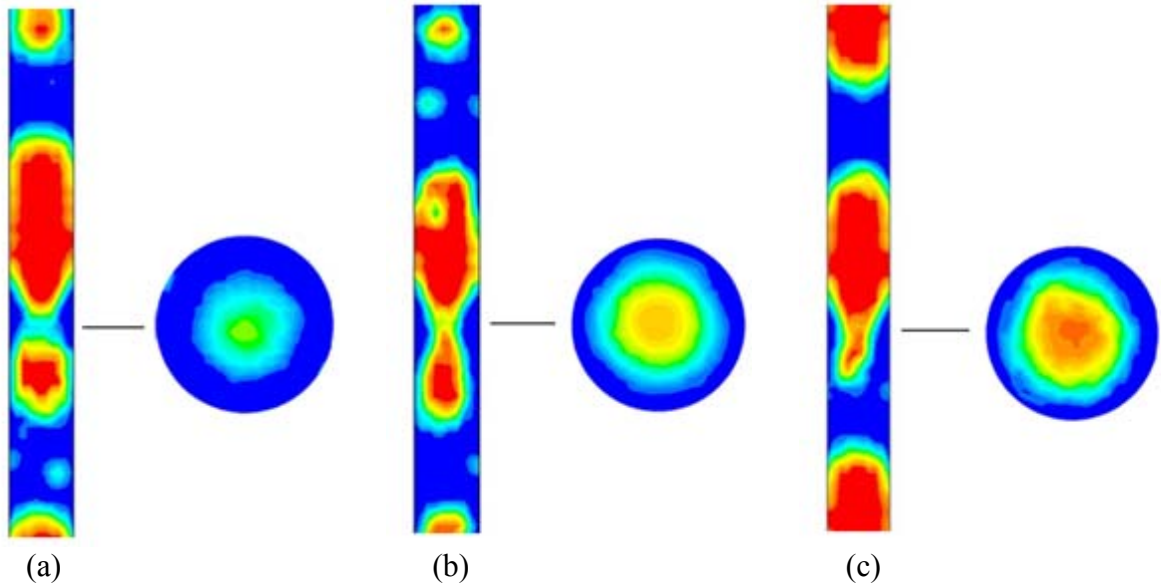

(a)            (b)            (c)

Figure 12 vapor bubbles combination

After volume fractions investigation, important approaches such spectral analysis, correlation dimension, autocorrelation function, Lyapunov exponent and phase space reconstruction were employed to investigate the chaotic state in the PHP. More details of the calculations in these approaches can be found in the references [26-29].

## 3.2. Spectral Analysis of Time Series

Temperature behavior of adiabatic wall of PHP is one of the most important factors to investigate chaos in the system. Thus, temperature time series analysis of the adiabatic wall is significant and crucial to analyze the chaotic state in a PHP. Recently there have been experimental studies that reported the existence of chaos in PHPs under some operating conditions. The approach in these studies is to analyze the time series of fluctuation of temperature of a specified location on the PHP tube wall (adiabatic section) by power spectrum calculated through Fast Fourier Transform (FFT) [21]. In this work, 10 points on the PHP are dedicated to measure the temperature of the adiabatic walls as shown in Fig. 1. To investigate and analyze the temperature time series, Power Spectrum Density (PSD) approach has been employed. The power spectrum of a time series describes the distribution of power into frequency components composing that signal. According to Fourier analysis any physical signal can be decomposed into a number of discrete frequencies or a spectrum of frequencies over a continuous range. The statistical average of a certain signal or sort of signal (including noise) as analyzed in terms of its frequency content is called its spectrum. Power spectrum density can be defined as [26]:

$$S_x(f) = \lim_{T \to \infty} E \left\{ \frac{1}{2T} \left| \int_{-T}^{T} x(t) e^{-j2\pi f t} dt \right|^2 \right\} \tag{14}$$

where x(t) is the random time signal and E is the energy of the signal. The power spectrum density is particularly useful for studying the oscillations of a system. There will be sharper or broader peaks at the dominant frequencies and at their integer multiples, the harmonics. Periodic or quasi-periodic signals show sharp spectral lines. Measurement noise adds a continuous floor to the spectrum. Thus in the spectrum, signal and noise are readily distinguished. Deterministic chaotic signals may also have sharp spectral lines but even in the absence of noise there will be a continuous part of the spectrum [27]. This is an immediate consequence of the exponentially decaying of autocorrelation function that will be shown in the next sections afterwards. The spectral analysis of the times series was carried out using Matlab codes in this paper.

Figure 13 shows the time series of the adiabatic wall temperature at point #16 under heating power of 90 W, condenser temperature of 20 ˚C, filling ratio of 45% and water as working fluid and its power spectrum density diagram. The spectral analysis of the time series in Fig. 13(b) indicates a dominant peak around frequency of 0.2 Hz. This signifies an intense periodic or quasi-periodic oscillation of temperature at this dominant frequency with PSD of 0.05 (W/Hz). PSD of oscillations at other frequencies are in an order to be neglected comparing with this dominant peak. Thus, the temperature behavior in Fig. 13(a) can be classified as periodic or quasi-periodic with frequency of 0.2 and PSD of 0.05 under mentioned operating conditions.

Figure 14 shows the time series of the adiabatic wall temperature at point #11 under heating power of 90 W, condenser temperature of 20 ˚C, filling ratio of 55% and ethanol as working fluid and its power spectrum density diagram. It can be seen in Fig. 14(b) that there are two dominant peaks in PSD diagram. Oscillations at frequencies of 0.12 Hz and 0.23 Hz have higher PSD of 0.05 and 0.03 (W/Hz) comparing oscillations at other frequencies. Existence of these two dominant peaks signifies periodic or quasi-periodic oscillations of temperature at these dominant frequencies. Thus, the temperature behavior in Fig. 14(a) cannot be classified as chaos under the mentioned operating conditions.

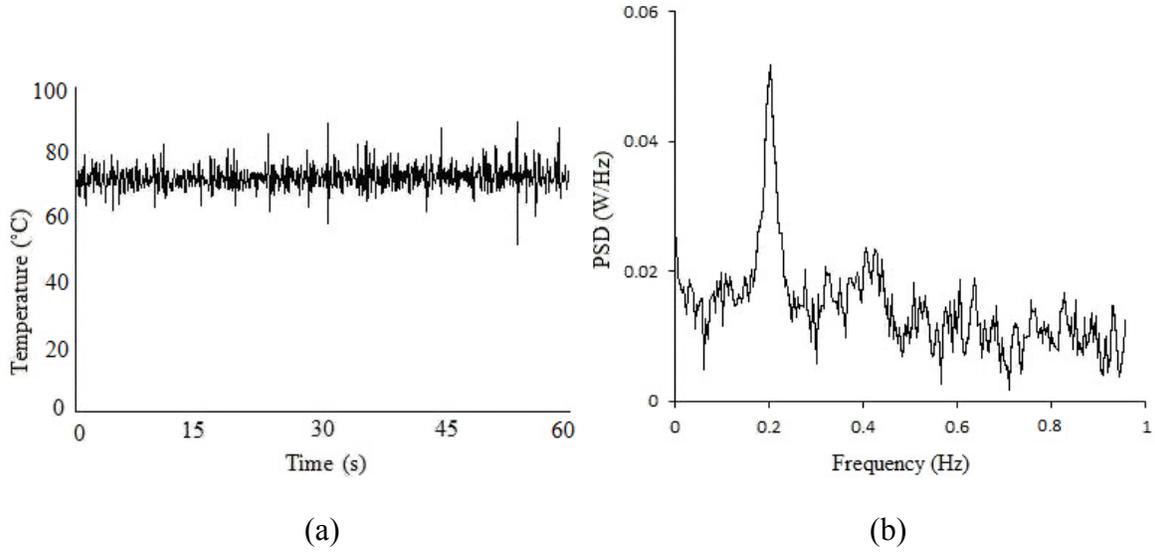

(a)                                  (b)

Figure 13 Time series of temperature (a) and PSD diagram (b) for point #16 under heating power of 90 W, condenser temperature of 20 ˚C, filling ratio of 45% and water as working fluid

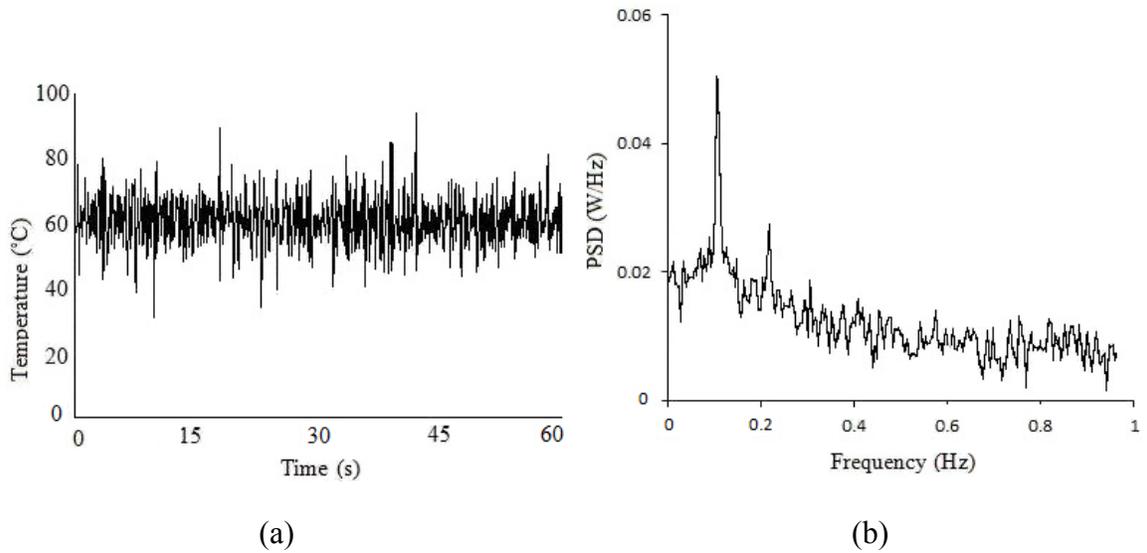

(a)                                  (b)

Figure 14 Time series of temperature (a) and PSD diagram (b) for point #11 under heating power of 90 W, condenser temperature of 20 ˚C, filling ratio of 55% and ethanol as working fluid

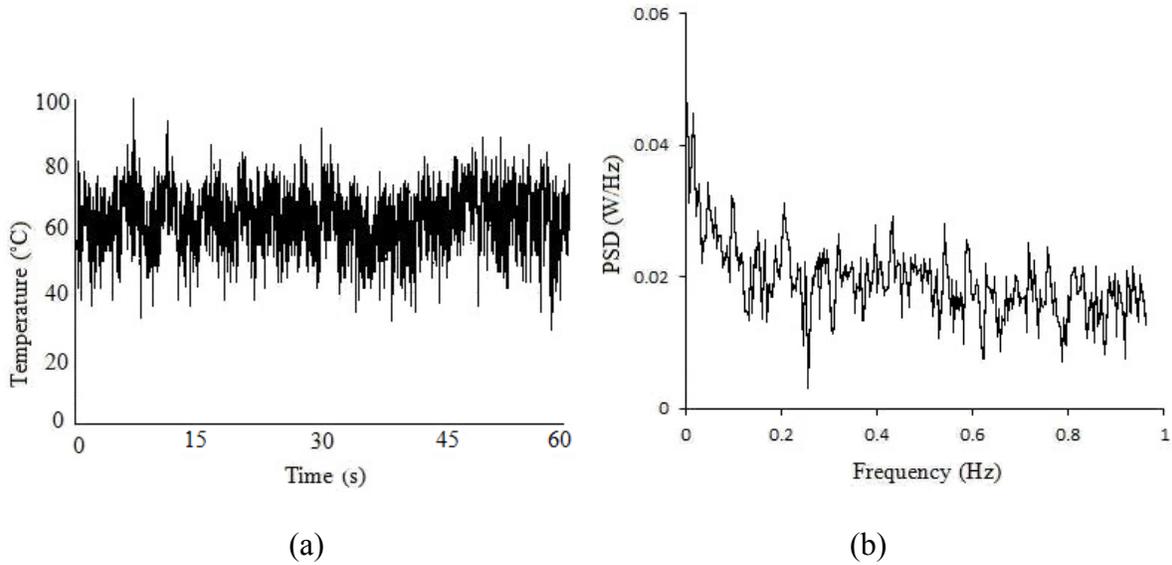

(a)                          (b)

Figure 15 Time series of temperature (a) and PSD diagram (b) for point #20 under heating power of 90 W, condenser temperature of 20 ˚C, filling ratio of 70% and water as working fluid

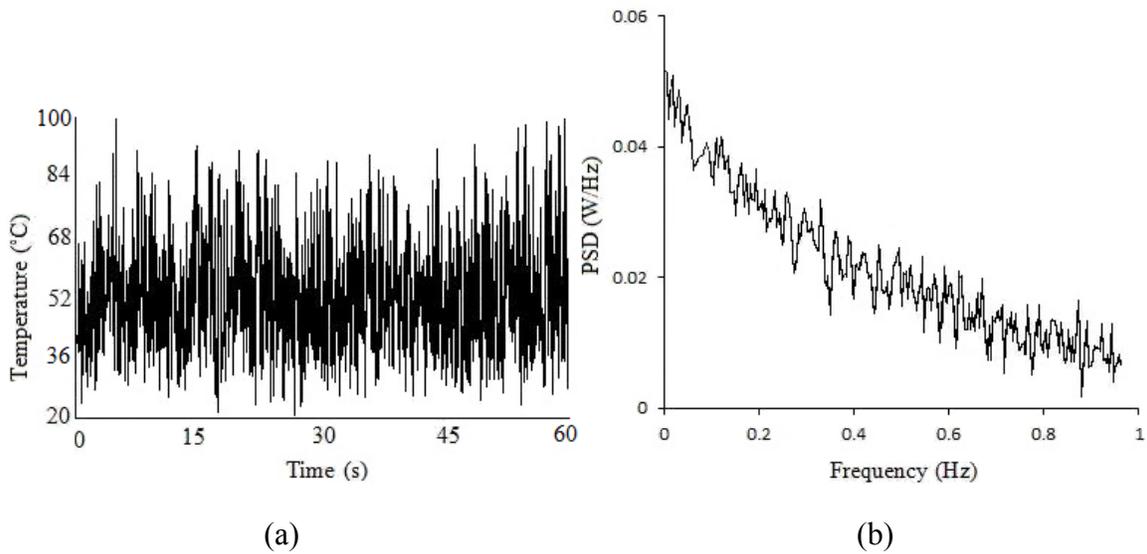

(a)                          (b)

Figure 16 Time series of temperature (a) and PSD diagram (b) for point #25 under heating power of 90 W, condenser temperature of 20 ˚C, filling ratio of 65% and ethanol as working fluid

Figure 15 shows the time series of the adiabatic wall temperature at point #20 under heating power of 90 W, condenser temperature of 20 ˚C, filling ratio of 70% and water as working fluid and its power spectrum density diagram. As shown in Fig. 15(b), there is not any significant and visible dominant peak in PSD diagram. At some frequencies, the PSD has higher value comparing other frequencies. But those higher values of PSD are not in an order to be considered as a dominant peak. Thus, the temperature oscillations are neither periodic nor quasi-periodic. In addition, by increasing the frequency, the power spectrum density decays in Fig. 15(b) as signature of chaotic behavior. Then, the temperature behavior in Fig. 15(a) can be classified as chaos under mentioned operating conditions.

Figure 16 shows the time series of the adiabatic wall temperature at point #25 under heating power of 90 W, condenser temperature of 20 °C, filling ratio of 65% and ethanol as working fluid and its power spectrum density diagram. It is evident there is not any dominant peak in PSD diagram at all. Besides, the power spectrum density decays by frequency increment. Absence of dominant peak in PSD diagram and its decay with respect to frequency indicate the chaotic state in the system. So, the temperature time series in Fig. 16(a) can strongly be classified as chaos under mentioned operating conditions.

As mentioned earlier, there are 10 points assigned for adiabatic wall temperature measurement (#1, #5, #6, #11, #15, #16, #20, #21 and #25). Numerical simulations and analytical investigations concluded that if any of these points had periodic or quasi-periodic behavior, rest of the points (in adiabatic section) had periodic or quasi-periodic behavior in terms of temperature time series. This conclusion was also applicable for chaotic behavior of the temperature time series. Then, in order to investigate the chaotic behavior in the entire PHP, only selecting one point (in adiabatic section) was sufficient.

### 3.3. Correlation Dimension

In chaos theory, the correlation dimension is a measure of the dimensionality of the space occupied by a set of data points, often referred to as a type of fractal dimension. The correlation dimension is well known in fractal geometry and it is used to calculate a fractal dimension from a time series. The time series is embedded in n-dimensional space which is done by forming vectors of length n. The embedding dimension of an attractor dataset is the dimension of its address space. In other words, it is the number of attributes of the attractor dataset. The attractor dataset can represent a spatial object that has a dimension lower than the space where it is embedded [29]. For example, a line has an intrinsic dimensionality one, regardless if it is in a higher dimensional space. The intrinsic dimension of an attractor dataset is the dimension of the spatial object represented by the attractor dataset, regardless of the space where it is embedded. The above definition of correlation dimension involves phase space vectors as the location of points on attractor. Thus, given a scalar time series, it is important to reconstruct an auxiliary phase space by an embedding procedure. Suppose that our attractor consists of N data points obtained numerically or experimentally, the correlation sum is defined as [27]:

$$C(r) = \lim_{N \to \infty} \left( \frac{number\ of\ pairs\ of\ points\ whitin\ a\ sphere\ of\ radius\ r}{N^2} \right) \quad (15)$$

In practice, we increase N until C(r) is independent of N. The correlation sum typically scales as

$$C(r) \approx \alpha\ r^{D_c} \quad (16)$$

where $D_c$ is the correlation dimension and r is chosen to be smaller than the size of attractor and larger than the smallest spacing between the points. In practice, we plot ln C(r) as a function of ln r and obtain the dimension $D_c$ from the slope of the curve.

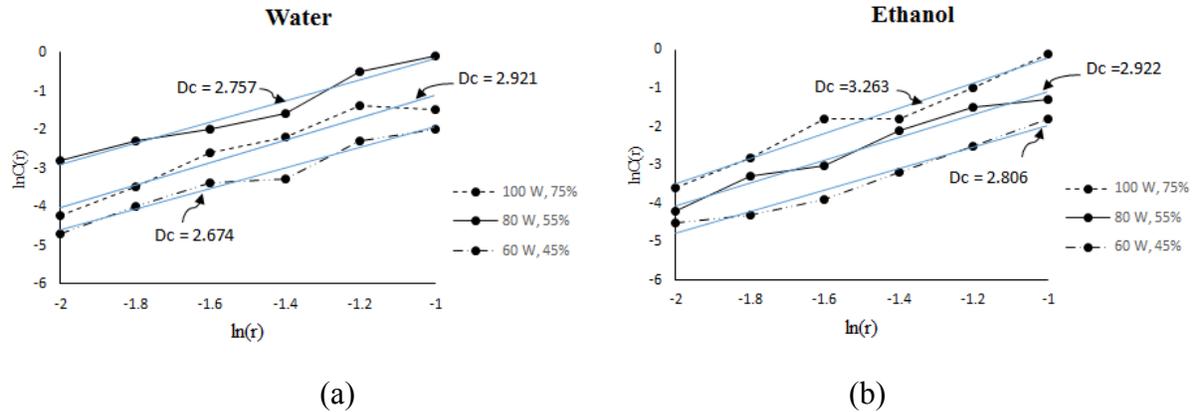

(a)            (b)

Figure 17 Correlation dimension values ($D_c$) with water (a) and ethanol (b) as working fluids

Figure 17 illustrates the correlation dimensions under evaporator heating powers of 60 W, 80 W and 100 W with filling ratios of 45%, 55% and 75%, respectively at condenser temperature of 20 °C for water and ethanol. Values of 2.674, 2.757 and 2.921 were obtained for water under mentioned operating conditions. In addition, values of 2.806, 2.922 and 3.263 were obtained for ethanol under the same operating conditions. It is evident for both water and ethanol that by increasing the evaporator heating power and filling ratio, the slope of the curves and correlation dimension increase. Besides, correlation dimension values for ethanol are higher than water under the same operating conditions. High values of correlation dimension refers to high frequency, small scale temperature oscillations, caused by miniature bubbles or short vapor plugs dynamically flowing in PHP tubes [26]. The lower correlation dimension corresponds to low frequency of temperature oscillations and large amplitude caused by large bubbles in the PHP. It is notable that value of correlation dimension indicates complexity of a system. Then by increasing the correlation dimension, the complexity of the PHP system increases.

### 3.4. Autocorrelation Function

Autocorrelation is the cross-correlation of a signal with itself at different points in time (that is what the cross stands for). Informally, it is the similarity between the observations as a function of the time lag between them. It is a mathematical tool for finding repeating patterns, such as the presence of a periodic signal obscured by noise, or identifying the missing fundamental frequency in a signal implied by its harmonic frequencies. It is often used in signal processing for analyzing functions or series of values, such as time domain signals. The autocorrelation of a random process describes the correlation between values of the process at different times, as a function of the two times or of the time lag. Often in time series it is required to compare the value observed at one time point to a value observed one or more time points earlier. Such prior values are known as lagged values. If there is some pattern in how the values of time series change from observations to observation, it would be so useful to analyze the time series. The correlation between the original time series values and the corresponding $\tau$-lagged values is called autocorrelation of order $\tau$. The Autocorrelation Function (ACF) provides the correlation between the serial correlation coefficients for consecutive lags. AFC helps to determine how quickly signals or process change with respect to time and whether a process has a periodic component. The autocorrelation function will have its largest value of AFC=1 at τ=0. Figure 18 illustrates the autocorrelation function (AFC) of time series at filling ratio of 60% and four heating powers of 50 W, 70 W, 90 W and 110 W. Autocorrelation function of the time series are computed as [27]:

$$ACF(\tau) = \frac{\sum_{i=1}^{N}(T_{i+\tau}-\bar{T})(T_i-\bar{T})}{\sum_{i=1}^{N}(T_i-\bar{T})^2} \qquad (17)$$

where $T_i$ and $T_{i+\tau}$ are the adiabatic wall temperature observations at the time domain, $\bar{T}$ is the overall mean temperature and $\tau$ is the lag.

Behavior of ACF indicates the prediction ability of the system [26]. The autocorrelation function will have its largest value of AFC=1 at τ=0. Figure 18 shows that the autocorrelation function decreases with time. Decreasing of the ACF shows that the prediction ability is finite as a signature of chaos. By increasing the evaporator heating power, ACF decreases for water and ethanol. In addition, linear behavior of the ACF changes gradually to exponential behavior by increasing the heating power. Although the same operating conditions were applied for the PHP with water and ethanol as working fluids, change in working fluid did not lead to any particular conclusion for ACF behavior.

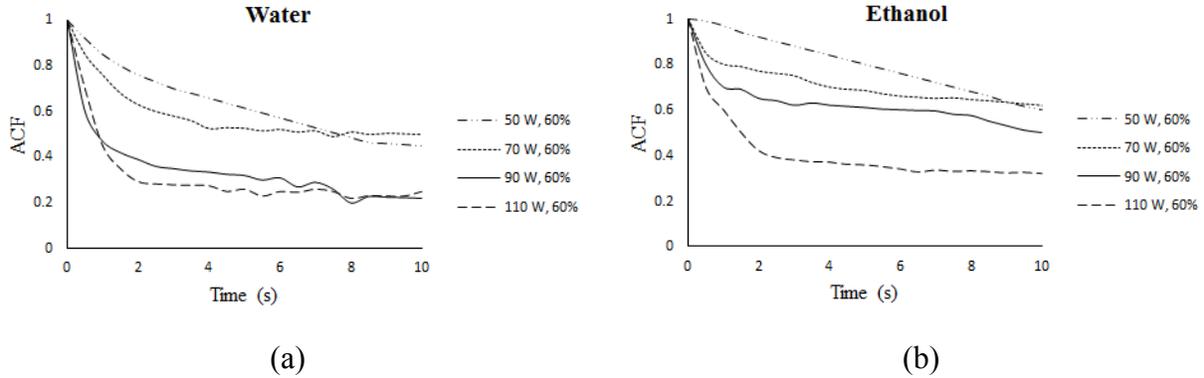

(a)          (b)

Figure 18 Autocorrelation function versus time with water (a) and ethanol (b) as working fluids

### 3.5. Lyapunov Exponent

Lyapunov exponents are dynamical quantities. Consider two points in a space, $X_0$ and $X_0 + \Delta x_0$, each of which will generate an orbit in that space using some equation or system of equations. These orbits can be thought of as parametric functions of a variable that is something like time. If we use one of the orbits a reference orbit, then the separation between the two orbits will also be a function of time. Because sensitive dependence can arise only in some portions of a system (like the logistic equation), this separation is also a function of the location of the initial value and has the form $\Delta x(X_0, t)$. In a system with attracting fixed points or attracting periodic points, $\Delta x(X_0, t)$ diminishes asymptotically with time. If a system is unstable, like pins balanced on their points, then the orbits diverge exponentially for a while, but eventually settle down. For chaotic points, the function $\Delta x(X_0, t)$ will behave erratically. It is thus useful to study the mean exponential rate of divergence of two initially close orbits using the formula. Let d(0) be the small separation between two arbitrary trajectories at time 0, and let d(t) be the separation between them at time t. Then for a chaotic system [29]:

$$d(t) \approx d(0) e^{\lambda t} \tag{18}$$

where $\lambda$ is the largest positive Lyapunov exponent. In the pulsating heat pipe, let $X_p$ be the position of the liquid plug relative to its initial position. Given some initial condition $X_0$, consider a nearby point $X_0+\delta$, where the initial separation $\delta_0$ is extremely small. Let $\delta_n$ be the separation after n steps. If $|\delta_n| \approx |\delta_0| e^{n\lambda}$, $\lambda$ will be the Lyapunov exponent. A positive $\lambda$ means an exponential divergence of nearby trajectories which is a signature of chaos. A negative $\lambda$ corresponds to stable fixed point and if $\lambda=0$, this means stable limit cycle. A more precise and computationally useful formula for $\lambda$ with the logistic map of $f(x)$ is derived by Strogatz [29]:

$$\lambda = \lim_{n\to\infty} \left[\frac{1}{n}\sum_{i=0}^{n-1} ln|f'(x_i)|\right] \tag{19}$$

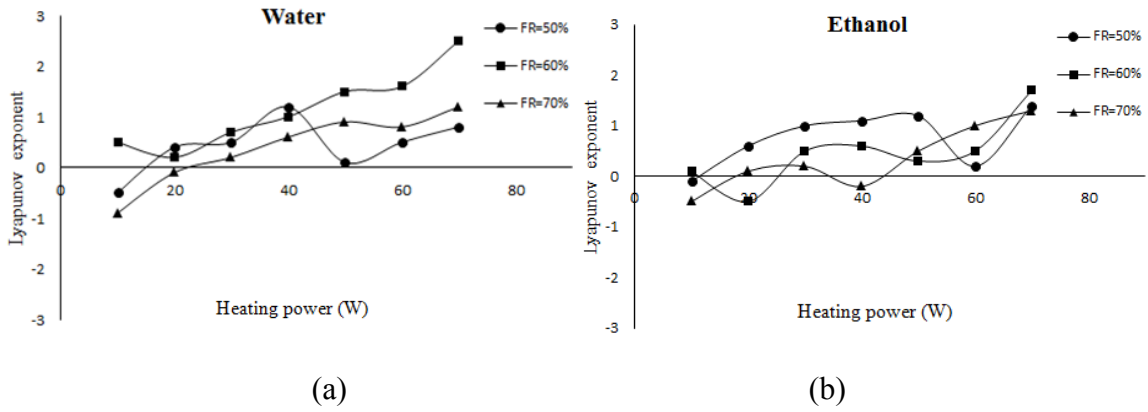

(a)          (b)

Figure 19 Lyapunov exponents versus evaporator heating power at different filling ratios with water (a) and ethanol (b) as working fluids

Figure 19 illustrates the Lyapunov exponents versus evaporator heating power at three filling ratio of 50%, 60% and 70% at condenser temperature of 20 ˚C with water and ethanol as working fluids. Mostly, Lyapunov exponent increases by increasing heating power in Fig. 19(a) with water as working fluid. Three exceptional points can be seen at heating powers of 20 W, 50 W and 60 W for filling ratios of 60 %, 50% and 70%. Negative Lyapunov exponents appeared at heating power of 10 W with FR of 70%, heating power of 10 W with FR of 50% and heating power of 20 W with FR of 70% indicating PHP is not in chaotic state under these operating conditions. Lyapunov exponent increases by increasing the heating power as well for the PHP with ethanol as working fluid as shown in Fig. 19(b). Exceptional points from this behavior exist under some operating conditions. Negative Lyapunov exponents appeared at heating power of 10 W with FR of 70%, heating power of 10 W with FR of 50%, heating power of 20 W with FR of 60% and heating power of 40 W with FR of 70% indicating PHP is not in chaotic state under these operating conditions. It is evident in Fig. 18 that range of Lyapunov exponent value for the PHP with water and ethanol as working fluids are similar. This indicates strong dependency of Lyapunov exponent to the structure and dimensions of the PHP which is the same for all operating conditions in current study.

## 3.6. Phase Space Reconstruction

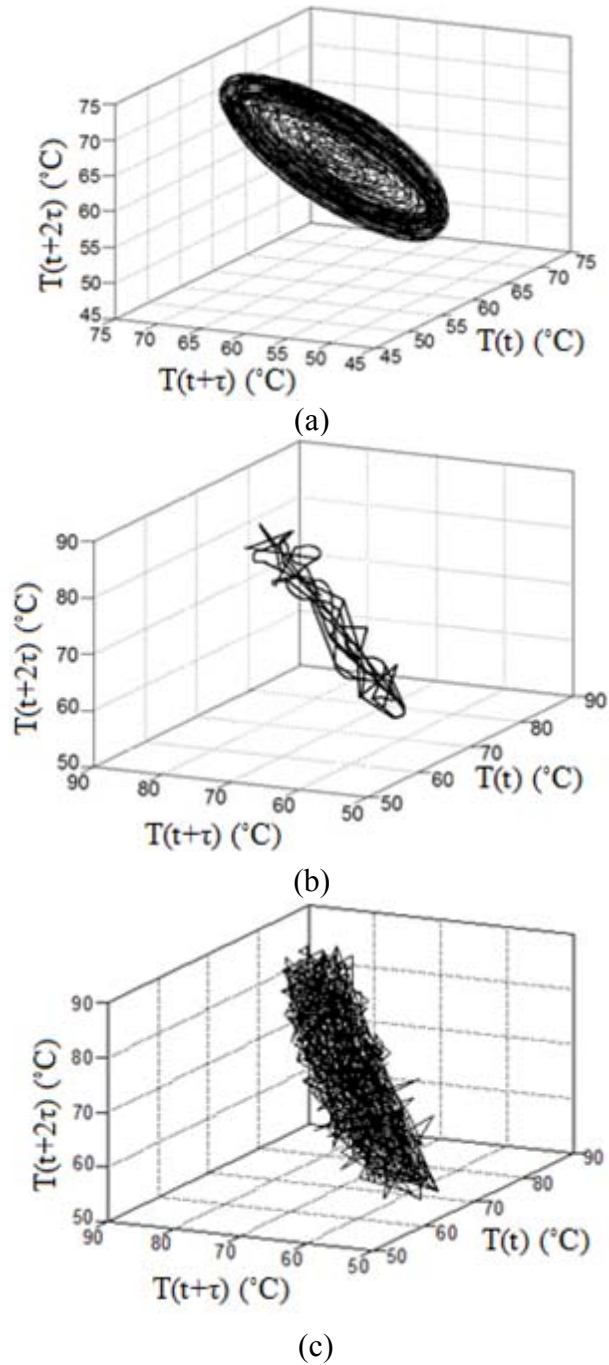

Figure 20 Reconstructed 3D attractor patterns under (a): heating power of 75 W, condenser temperature of 20 °C, filling ratio of 55% and ethanol as working fluid (b): heating power of 40 W, condenser temperature of 20 °C, filling ratio of 60% and water as working fluid (c): heating power of 80 W, condenser temperature of 20 °C, filling ratio of 65% and ethanol as working fluid

To reconstruct a useful version of the internal dynamics for a given time series from a sensor on a single state variable $x_i(t)$ in a n-dimensional dynamical system, delay-coordinate embedding can be used. If the embedding is performed correctly, the involved theorems guarantee that the reconstructed dynamics is topologically identical to the true dynamics of the system, and therefore that the dynamical invariants are also identical. This is an extremely powerful correspondence; it implies that conclusions drawn from the embedded or reconstruction-space dynamics are also true of the real unmeasured dynamics. This implies, for example, that one can reconstruct the dynamics of the earth's weather simply by setting a thermometer on a windowsill [28]. Considering a data set comprised of samples $x_i(t)$ of a signal state variable $x_i$ in a n-dimensional system, measured once every $\Delta t$ seconds. To embed such a data set, we construct r-dimensional reconstruction-space vectors *T(t)* from r time-delayed samples of the $x_i(t)$, such that

*T(t)=[$x_i(t)$, $x_i(t+\tau)$, $x_i(t+2\tau)$,…, $x_i(t+(m-1)\tau)$]*                                    (20)

In a pulsating heat pipe, the time series is usually a sequence of temperature values which depends on the current phase space, taken at different points in time. It is important that given enough dimensions (r) and the right delay ($\tau$), the reconstruction-space dynamics and the true, unobserved state-space dynamics are topologically identical. More, formally, the reconstruction-space and state-space trajectories are guaranteed to be equivalent if $r \geq 2n+1$.

Regarding above explanations, the attractor reconstructions were carried out. It is evident in Fig. 20(a) the attractor pattern is like an O-ring structure which is mostly for periodic or quasi-periodic systems. Then, the PHP behavior under heating power of 75 W, condenser temperature of 20 °C, filling ratio of 55% and ethanol as working fluid is not in a chaotic state. Figure 20(b) illustrates the attractor pattern under heating power of 40 W, condenser temperature of 20 °C, filling ratio of 60% and water as working fluid. It can be seen that the attractor pattern at this low heating power is narrow, slim and elongated, corresponding to low dimension characteristic of the temperature oscillation and chaotic properties of PHP. Figure 20(c) illustrates the attractor pattern under heating power of 80 W, condenser temperature of 20 °C, filling ratio of 65% and ethanol as working fluid. At this higher heating power, the attractor pattern is more complex. In addition, it has slight wider and more scattered distribution than the attractor pattern in Fig. 20(b), corresponding to a higher dimension characteristic of the temperature oscillation and chaotic properties of PHP. Figures 20(b) and 20(c) are strong evidence of deterministic chaotic systems under mentioned operating conditions.

### 3.7. Thermal Performance

In current study, effects of filling ratio and working fluid on the thermal performance of the PHP have been investigated. Filling ratios of 35%, 45%, 55%, 65% and 75% for water and ethanol have been tested under different heating powers and condenser temperature of 20 °C. For the traditional heat pipes, the filling ratio range of working fluid is lower comparing to the PHPs. Large amount of working fluid inside the channels causes liquid blockage and reduces pressure difference between heating and cooling sections which can stop the oscillation movement. But for the PHPs a filling ratio that is higher than that of traditional heat pipes is required. The overall thermal resistance of a PHP is defined as the difference average temperatures between evaporator and condenser divided by the heating power. For evaporator average temperature, 15 points (#2, #3,

#4, #7, #8, #9, #12, #13, #14, #17, #18, #18, #22, #23 and #24) were used to calculate the average temperature. Since condenser has constant temperature, calculation is not necessary to get the average temperature which is equal to that constant value. Equation (21) defines the thermal resistance of a PHP.

$$R = \frac{Te-Tc}{Q} \qquad (21)$$

where $T_e$ is the evaporator average temperature, $T_c$ is the condenser average temperature and Q is the evaporator heating power.

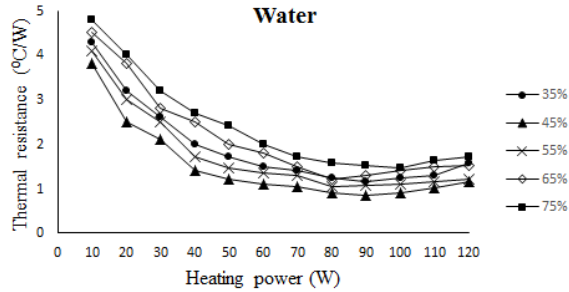

Figure 21 Thermal resistance versus heating power at different filling ratios and water as working fluid

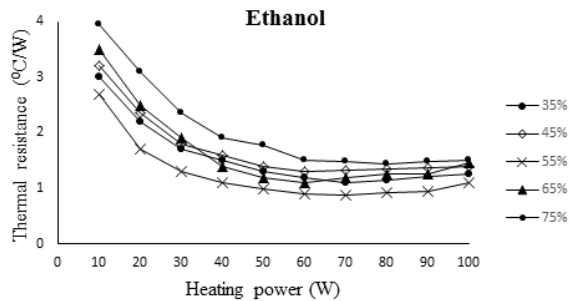

Figure 22 Thermal resistance versus heating power at different filling ratios and ethanol as working fluid

Figure 21 illustrates the thermal resistance variation with respect to evaporator heating power at different filling ratios and water as working fluid. A general behavior of thermal resistance can be seen in Fig. 21. At all filling ratios, by increasing the heating power, the thermal resistance decreases initially to reach a minimum value as an optimal point. Thermal resistance increases by increasing the heating power afterwards. This optimal point is important since at related operating conditions, the pulsating heat pipe transfers the maximum heat flux with a minimum temperature difference between the evaporator and condenser leads to a better thermal performance of the PHP. The minimum thermal resistance occurs at different heating powers for different filling ratios. It is obvious in Fig. 20 that thermal resistance has lower values at filling ratio of 45% comparing with other filling ratios. The minimum thermal resistance of 0.85 ˚C/W happens at heating power of 90 W for this filling ratio.

As shown in Fig. 22, the same conclusion of general behavior and optimal point of thermal resistance is applicable for ethanol as well. Thermal resistance has lower values at filling ratio of 55%. The minimum thermal resistance of 0.88 ˚C/W happens at heating power of 70 W for this filling ratio.

## 4. Conclusion

Numerical simulations have been carried out to investigate the chaotic and thermal behaviors of a 3D closed-loop pulsating heat pipe. Heat flux and constant temperature boundary conditions were applied for evaporator and condenser respectively. Water and ethanol were used as working fluids. Volume fractions of liquid and vapor were obtained and analyzed under different operating conditions. Comparison of thermal resistance behavior versus heating power between simulation results of current study and experimental results of Shafii et al. [13] confirmed a good agreement between simulation and experimental results. The following conclusions can be drawn from this work:

- Spectral analysis of temperature time series using Power Spectrum Density showed existence of dominant peak in PSD diagram indicates periodic or quasi-periodic behavior in temperature oscillations at particular frequencies.
- Absence of dominant peak in PSD diagram and its decay with respect to time was a signature of chaotic state under some operating conditions.
- It was found for both water and ethanol as working fluids by increasing the evaporator heating power and filling ratio, correlation dimension increases.
- Correlation dimension values for ethanol were higher than water under the same operating conditions. High values of correlation dimension referred to high frequency, small scale temperature oscillations, caused by miniature bubbles or short vapor plugs dynamically flowing in PHP tubes.
- Decay of autocorrelation function with respect to time indicated finite prediction ability of the system. Change in working fluid did not lead to any particular conclusion for ACF behavior.
- Similar range of Lyapunov exponent value for the PHP with water and ethanol as working fluids indicated strong dependency of Lyapunov exponent to the structure and dimensions of the PHP.
- An O-ring structure pattern was obtained for reconstructed 3D attractor at periodic or quasi-periodic behavior of temperature oscillations.
- Optimal points were found in case of thermal performance with minimum thermal resistance of 0.85 ˚C/W at heating power of 90 W and filling ratio of 45% for water as working fluid and minimum thermal resistance of 0.88 ˚C/W at heating power of 70 W and filling ratio of 55% for ethanol as working fluid.

**Nomenclature**

$\boldsymbol{F}$  Body force (kg·m/s$^2$)
$\boldsymbol{g}$  Gravity acceleration (m/s$^2$)
k  Thermal Conductivity (W/m.K)
$\boldsymbol{n}$  Surface normal vector
T  Temperature (°C)
$\alpha_q$  Void Fraction of phase q
$\boldsymbol{v}_q$  Velocity of phase q (m/s)
$\rho_q$  Density of phase q (kg/m$^3$)
$\mu$  Dynamic viscosity (kg/m.s)
$\sigma_{ij}$  Surface tension (kg/s$^2$)
$\theta_w$  Contact angle at the wall
$f$  Frequency (Hz)
λ  Lyapunov exponent